\documentstyle[preprint,aps,epsf]{revtex}

\def\beq{\begin{equation}}
\def\eeq{\end{equation}}
\def\6{\langle}
\def\9{\rangle}

\begin{document}
\draft

\title{Quantum cryptography with 3-state systems}

\author{Helle Bechmann-Pasquinucci$^1$ and Asher Peres$^2$}
\address{$^1$Group of Applied Physics, University of Geneva, CH-1211,
Geneva 4, Switzerland\\
$^2$Department of Physics, Technion---Israel Institute of Technology,
32000 Haifa, Israel}

\maketitle
\begin{abstract}
We consider quantum cryptographic schemes where the carriers of
information are 3-state particles. One protocol uses four mutually
unbiased bases and appears to provide better security than
obtainable with 2-state carriers. Another possible method allows quantum
states to belong to more  than one basis. The security is not better,
but many curious features arise. \end{abstract}

\bigskip
\pacs{PACS numbers: 03.67.-a, 03.67.Dd, 03.65.Bz}

When Samuel Morse invented the telegraph, he devised for it an alphabet
consisting of three symbols: dash, dot, and space. More modern
communication methods use binary signals, conventionally called 0 and~1.
Information theory, whose initial goal was to improve communication
efficiency, naturally introduced binary digits (bits) as its accounting
units. However, the theory can easily be reformulated in terms of
ternary digits (trits) 0, 1,~2, or larger sets of symbols~\cite{moscow}.
For example, instead of bytes (sets of 8 bits) representing 256 ordinary
characters, we would have ``trytes'' (5 trits) for 243 characters. An
ordinary text would thus be encoded into a string of trits. If we wish
to encrypt the latter, this can be done by adding to it (modulo~3) a
{\it random\/} string, called {\it key\/}, known only to legitimate
users. Decrypting is then performed by subtracting that key (modulo~3).

The aim of quantum cryptography~\cite{bb84} is to generate a secret key
by using quantum carriers for the initial communication between distant
parties (conventionally called Alice and Bob). The simplest methods use
2-state systems, such as polarized photons. Ortho\-gonal states
represent bit values 0 and~1. One may use either two~\cite{bb84} or
three \cite{dagmar,6st} orthogonal bases, chosen in such a way that any
basis vectors $|e_j\9$ and $|e_\mu\9$ belonging to different bases
satisfy $|\6e_j,e_\mu\9|^2=1/2$. Such bases are called {\it mutually
unbiased\/} \cite{ivan,wkw}. As a consequence, if an eavesdropper (Eve)
uses the wrong basis, she gets no information at all and causes maximal
disturbance (error rate 1/2) to the transmission, thereby revealing her
presence.

In this Letter, we consider 3-state systems as the quantum carriers for
cryptographic key distribution. For example, one may use
``biphotons''~\cite{biphoton}, namely photon pairs in symmetric Fock
states $|0,2\9$, $|2,0\9$, and $|1,1\9$. Biphotons can easily be
produced with present technology, and detecting arbitrary linear
combinations of them will probably be possible soon. (Another
possibility would be to use four states of a pair of
photons~\cite{tittel}, but here we consider only 3-state systems.)

Following the method of refs.~\cite{dagmar,6st}, we introduce four
mutually unbiased bases. Let $|\alpha\9$, $|\beta\9$, and $|\gamma\9$ be
the unit vectors of one of the bases. Another basis is obtained by a
discrete Fourier transform,

\beq\begin{array}{l}
|\alpha'\9=(|\alpha\9+|\beta\9+|\gamma\9)/\sqrt{3},\\ 
|\beta'\9=
(|\alpha\9+e^{2\pi i/3}|\beta\9+e^{-2\pi i/3}|\gamma\9)/\sqrt{3},\\
|\gamma'\9=
(|\alpha\9+e^{-2\pi i/3}|\beta\9+e^{2\pi i/3}|\gamma\9)/\sqrt{3}.
\end{array}\eeq
The two other bases can be taken as

\beq (e^{2\pi i/3}|\alpha\9+|\beta\9+|\gamma\9)/\sqrt{3}\qquad
 \mbox{and cyclic perm.},\eeq
and

\beq (e^{-2\pi i/3}|\alpha\9+|\beta\9+|\gamma\9)/\sqrt{3}\qquad
 \mbox{and cyclic perm.}.\eeq
Any basis vectors $|e_j\9$ and $|e_\mu\9$ belonging to different bases
now satisfy $|\6e_j,e_\mu\9|^2=1/3$.

The protocol for establishing a secret key is the usual one. Alice
randomly chooses one of the 12 vectors and sends to Bob a signal whose
quantum state is represented by that vector. Bob randomly chooses one of
the four bases and ``measures'' the signal (that is, Bob tests whether
the signal is one of the basis vectors). Having done that, Bob publicly
reveals which basis he chose, but not the result he obtained. Alice then
reveals whether her vector belongs to that basis. If it does, Alice and
Bob share the knowledge of one trit. If it does not, that transmission
was useless. This procedure is repeated until Alice and Bob have
obtained a long enough key. They will then have to sacrifice some of the
trits for error correction and privacy amplification~\cite{privacy} (we
shall not discuss these points, which are the same as in all
cryptographic protocols, except that we have to use trits instead of
bits, and therefore parity checks become triality checks, that is, sums
modulo~3).

Consider the simplest eavesdropping strategy: Eve intercepts a particle,
measures it, and resends to Bob the state that she found. In 3/4 of the
cases, she uses a wrong basis, gets no information, and causes maximal
disturbance to the transmission: Bob's error rate (that is, the
probability of a wrong identification of the trit value) is 2/3. On the
average, over all transmissions, Eve gets $I_E=1/4$ of a trit and Bob's
error rate is $E_B=1/2$. (It is natural to measure Eve's information in
trits, since Bob gets one trit for each successful transmission.) These
results may be compared to those obtained by using 2-state systems. With
only two bases as in ref.~\cite{bb84} and with the same simple
eavesdropping strategy, Eve learns on the average 1/2 of a bit for each
transmitted bit, and Bob's error rate is 1/4. If we use three bases as
in \cite{dagmar,6st}, these numbers become 1/3. Thus, with the present
method, Eve learns a smaller fraction of the information and causes a
larger disturbance. It is likely that this is also true in presence of
more sophisticated eavesdropping strategies, such as using an ancilla to
gently probe the transmission without completely disrupting it. When
people seek Eve's ``optimal'' eavesdropping strategy~\cite{opti}, their
criterion usually is the maximal value of $I_E/E_B$.

Do the above results mean that using 3-state systems improves the
cryptographic security? The answer depends on which aim we seek to
achieve. If Alice and Bob simply wish to be warned that an eavesdropper
is active, and in that case they will use another communication channel,
then obviously the highest possible ratio $E_B/I_E$ is desirable. Eve can
at most conceal her presence by intercepting only a small fraction $x$
of the transmissions, such that $xE_B$ is less than the natural error
rate, but then $I_E$ is reduced by the same factor, and Eve's illicit
information can be eliminated by classical privacy
amplification~\cite{privacy}.

However it may be that Alice and Bob have no alternative channel to use
and privacy amplification is their only possibility of fighting the
eavesdropper. In that case, it is known \cite{CK,ehpp} that secure
communication can in principle be achieved if Bob's mutual information
with Alice, $I_B$, is larger than Eve's $I_E$. Note that even if Bob and
Eve have the same error rate, as in one of the above examples,
$I_E>I_B$. The reason is that Eve knows whether Alice and Bob used the
same basis, and therefore which ones of her data are correct and which
ones are worthless. On the other hand, Bob can only compare with Alice a
subset of data, so as to measure his mean error rate $E_B$, and from the
latter deduce the Shannon entropy of his string. For 2-state systems,
assuming all bit values equally probable, he obtains

\beq I_B=1+(1-E_B)\log_2(1-E_B)+E_B\log_2E_B, \label{ib1}\eeq
and likewise for 3-state systems,

\beq I_B=1+(1-E_B)\log_3(1-E_B)+E_B\log_3(E_B/2). \label{ib2}\eeq
Numerical results, in bits and trits respectively, will be given in
Table II at the end of this Letter, together with those for two other
cryptographic protocols, discussed below.

New types of cryptographic protocols may indeed be devised if the
Hilbert space has more than two dimensions. The reason is that a basis
vector may now belong to several bases. In that case, it is natural to
assume that each vector represents a definite trit (0, 1, or 2), which
is the same in all the bases to which that vector belongs~\cite{diff}.
An example is given in the table below, where vectors are labelled
green, red, and blue, for later convenience.

\begin{quote}TABLE I. \ Components of 21 unnormalized vectors. The
symbols \=1\ and \=2\\ stand for $-1$ and $-2$, respectively. Orthogonal
vectors have different colors.\end{quote}
\begin{center}\begin{tabular}{lccccccccc}\hline
green& \ \ \ &001&101&0\=11&1\=11& \ \ &1\=12&112&2\=11\\
red&&100&110&10\=1&11\=1&&21\=1&211&12\=1\\
blue&&010&011&\=110&\=111&&\=121&121&\=112\\ \hline\end{tabular}
\end{center}\bigskip

Although this new algorithm does not improve transmission security (as
shown below), it has many fascinating aspects and leads to new insights
into quantum information theory. The 12 vectors in the first four
columns of Table~I are shown in Fig.~1, as dots on the faces of a cube,
in a way similar to the graphical representation of a Kochen-Specker
uncolorable set~\cite{qt}. In the present case, the tricolor analogue of
the Kochen-Specker theorem requires only 13 rays for its proof, because
ray (111) is orthogonal to all the rays in the third column, which have
three different colors. These 12 vectors form 13 bases, but only four
bases are complete. The nine others bases have only two vectors each and
have to be completed by nine new vectors, listed in the last three
columns of the table. To display these nine vectors on Fig.~1, their
integer components should be divided by 2. The corresponding dots are
then located at the centers of various squares on the faces of the
cube.

The cryptographic protocol is the same as before, but now Alice has 21
vectors to choose from, and Bob has a choice of 13 bases. The essential
difference is that these bases are not mutually unbiased, so that if Eve
chooses a different basis (which happens 12/13 of the time), she still
gets {\it at least\/} probabilistic information on Alice's vector.  It
may also happen that Eve's basis is different from Bob's, but both bases
contain the vector found by Eve. In that case, when Bob announces his
basis and Alice confirms it, Eve can infer that she got the correct
state and caused no error.

Let us analyze what happens for each successful transmission, that is,
when Alice's vector $|e_j\9$ is one of those in the basis announced by
Bob. Suppose that in her eavesdropping attempt, Eve obtains a state
$|e_\mu\9$. This happens with probability $P_{\mu j}=|\6e_j,e_\mu\9|^2$.
This is also the probability that Bob gets the correct $|e_j\9$ when Eve
resends to him $|e_\mu\9$. On the average over all Alice's $|e_j\9$ and
all Eve's choices of a basis, the probability that Bob gets a correct
result is

\beq C=
 \sum_{j=1}^{21}\sum_{\mu=1}^{21} M_\mu\,(P_{\mu j})^2/(21\times13),\eeq
where $M_\mu$ is the number of bases to which $|e_\mu\9$ belongs (namely
$M_\mu=2$ for the vectors in the first and third columns of Table~I, 
$M_\mu=3$ for those of the second and fourth columns, and $M_\mu=1$ for
the rest). Bob's mean error probability is $E_B=1-C=0.385022$.

To evaluate Eve's gain of information $I_E$, we note that when Alice
confirms the basis chosen by Bob, Eve is left with a choice of three
vectors having equal prior probabilities, $p_j=1/3$. The initial Shannon
entropy is $H_i=1$~trit, and the prior probability for Eve's result
$\mu$ is

\beq q_\mu=\sum_{j=0}^2 P_{\mu j}\,p_j=1/3. \eeq
It then follows from Bayes's theorem that the likelihood (posterior
probability) of signal $j$ is (see ref.~\cite{qt}, page 282)

\beq Q_{j\mu}=P_{\mu j}\,p_j/q_\mu=P_{\mu j}. \eeq
The new Shannon entropy, following result $\mu$, is

\beq H_f=-\sum_{j=0}^2 Q_{j\mu}\,\log_3 Q_{j\mu}. \eeq
Eve's information gain is obtained by averaging $H_f$ over all results
$\mu$, all Eve's bases, and all Bob's bases. The final result is

\begin{eqnarray} I_E&=&H_i-\6H_f\9,\\
 &=&1+\sum_{j=1}^{21}\sum_{\mu=1}^{21} M_j\,M_\mu\,
  P_{\mu j}\,\log_3 P_{\mu j}\;/(3\times13^2).\end{eqnarray}
Table II lists the relevant data for intercept-and-resend eavesdropping
(IRE) on all the above cryptographic protocols.

\begin{quote}TABLE II. Result of IRE on various cryptographic protocols:
Eve's information; Bob's information and error rate for a single IRE
event; and fraction of eavesdropped transmissions needed to make both
informations equal to each other. \end{quote}
\begin{center}\begin{tabular}{cccccccccccc}\hline
units&&bases&&vectors&$I_E$&&$I_B$&&$E_B$&&$x$\\ \hline
bits&&2&&4&0.500000&&0.188722&&0.250000&&0.68214\\
bits&&3&&6&0.333333&&0.081710&&0.333333&&0.68128\\
trits&&4&&12&0.250000&&0.053605&&0.500000&&0.71770\\
trits&&13&&12&0.575142&&0.143418&&0.391738&&0.51007\\
trits&&13&&21&0.442765&&0.150431&&0.385022&&0.68994\\
\hline\end{tabular} \end{center}\bigskip

We also investigated the possibility that Alice uses only the 12 vectors
in the first four columns of Table~I (those represented by the dots in
Fig.~1). The IRE results are also listed in Table~II. However, it is
interesting that in this case, Eve can get some information without
performing any active eavesdropping and without causing any error, just
by passively listening and waiting for Alice to confirm Bob's choice of
an incomplete basis. Eve then learns that one of the three trit values
is eliminated. On the average, she gets information

\beq I_E=(9/13)\,(1+\log_3 2)=0.255510\quad\mbox{trit}. \eeq

Finally, let us investigate what happens if Eve eavesdrops only on a
fraction $x$ of the particles sent by Alice. In that case, both $I_E$
and $E_B$ are multiplied by $x$, and $I_B$ is still given by
Eqs.~(\ref{ib1}) and~(\ref{ib2}), with $E_B$ replaced by $xE_B$ on the
right hand side. The results are displayed in Fig.~2, which also shows
the security domain $I_B\ge I_E$, assuming standard error correction and
privacy amplification~\cite{privacy}. The values of $x$ for which
$I_B=I_E$ are listed in the last column of Table~II. We see that the use
of four mutually unbiased bases for 3-state particles requires the
highest value of $x$ to breach the security. Moreover, for any given
value of $I_B$, this protocol is the one that gives the lowest value of
$I_E$. It thus appears that this method is the one giving the best
results against IRE attacks. It is likely to also be the best for more
sophisticated eavesdropping strategies, but this problem lies beyond the
scope of the present Letter.

\bigskip We thank Nicolas Gisin for helpful comments on cryptographic
security, and Daniel Terno for bringing ref.~\cite{moscow} to our
attention. H.B.-P. was supported by the Danish National Science Research
Council (grant no.\ 9601645) and also acknowledges support from the
European IST project EQUIP. \quad A.P. was supported by the Gerard Swope
Fund and the Fund for Encouragement of Research.

\newpage

\vspace*{\fill}
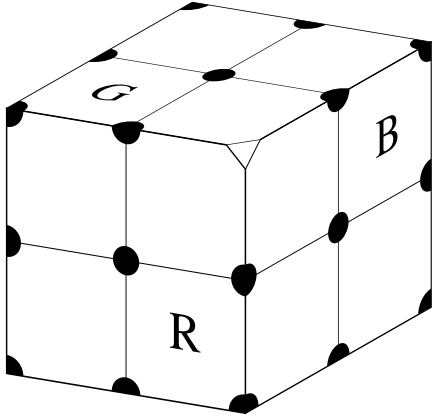
\noindent FIG. 1. \ Twelve vectors are obtained by connecting the center
of the cube to the various dots on its faces (diametrically opposite
dots represent the same vector). The four dots at the vertices of the
squares labelled G, R, and B, are green, red and blue, respectively. The
truncated vertex corresponds to the uncolorable ray (111).


\newpage
\vspace*{\fill}
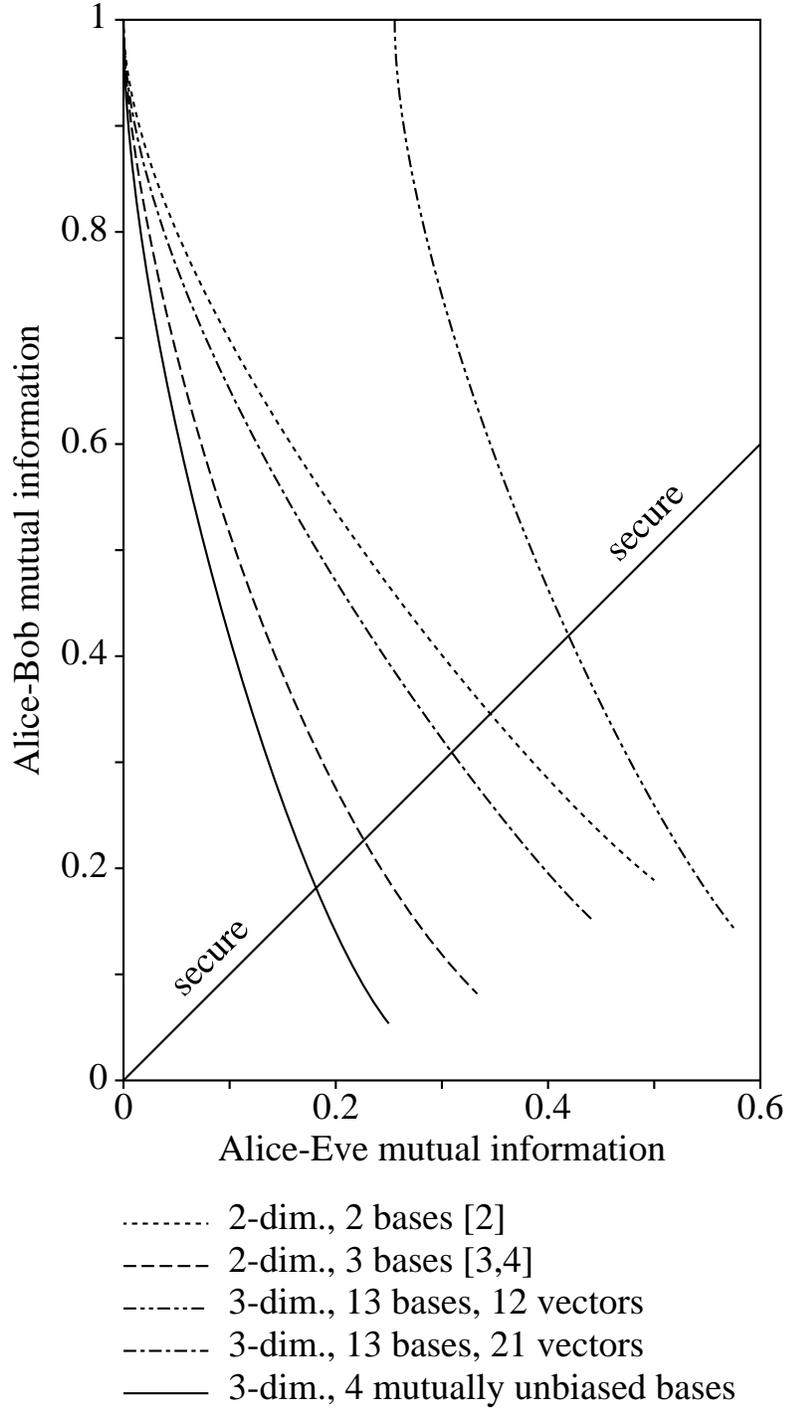
\noindent FIG. 2. \ Mutual informations for the various protocols listed
in Table~II, when the fraction of intercepted particles is $0<x<1$. For
the case of 13 bases and 12 vectors, it is assumed that in the remaining
fraction $(1-x)$, Eve performs passive eavesdropping on incomplete
bases. The data are given in bits for 2-dimensional systems, and trits
for 3-dimensional ones.


\begin{thebibliography}{99} \itemsep 0pt

\bibitem{moscow}A computer with ternary logic was built at Moscow State
University; this is reported in the Russian translation of D.~Knuth,
{\it The Art of Computer Programming\/} (Nauka, Moscow, 1976) vol.~1,
p.~156 (comment of the translator).

\bibitem{bb84}C.H. Bennett and G. Brassard, in {\it Proceedings of IEEE
International Conference on Computers, Systems and Signal Processing,
Bangalore, India\/} (IEEE, New York, 1984) p.~175.

\bibitem{dagmar}D. Bru\ss, Phys. Rev. Lett. {\bf81}, 3018 (1998).
 
\bibitem{6st}H. Bechmann-Pasquinucci and N. Gisin, Phys. Rev. A {\bf59},
4238 (1999).

\bibitem{ivan}I. D. Ivanovi\'c, J. Phys. A: Math. Gen. {\bf14}, 3241
(1981).

\bibitem{wkw}W. K. Wootters, Found. Phys. {\bf16}, 391 (1986).

\bibitem{biphoton}A. V. Burlakov, M. V. Chekhova, O. A. Karabutova, 
D. N. Klyshko, and S. P. Kulik, Phys. Rev. A {\bf60}, R4209 (1999).

\bibitem{tittel}H. Bechmann-Pasquinucci and W. Tittel,
quant-ph/9910095.

\bibitem{privacy}C.~H.~Bennett, F.~Bessette, G.~Brassard, L.~Salvail,
and J.~Smolin, J. Crypto. {\bf5}, 3 (1992).

\bibitem{opti}C. A. Fuchs, N. Gisin, R. B. Griffiths, C.-S. Niu, and A.
Peres, Phys. Rev. A {\bf56}, 1163 (1997).

\bibitem{CK}I.~Csisz\'ar and J.~K\"orner, IEEE Trans. on Information
Theory {\bf24}, 339 (1978).

\bibitem{ehpp}A. K. Ekert, B. Huttner, G. M. Palma, and A. Peres,
Phys. Rev. A {\bf50}, 1047 (1994).

\bibitem{diff}We also considered the possibility that a vector has
different values in different bases. This brings no improvement, since
Eve will learn which basis was actually used by Bob.

\bibitem{qt}A. Peres, {\it Quantum Theory: Concepts and Methods\/}
(Kluwer Academic Publishers, Dordrecht, 1995) p.~198.

\end{thebibliography}
\end{document}